\documentclass[apjl]{emulateapj}
\usepackage{graphicx}
\usepackage[]{natbib}
\usepackage{hyperref}

\def \ergsec{\hbox{erg s$^{-1}$}}

\def \phcmsec{\hbox{photons cm$^{-2}$ s$^{-1}$}}
\def \ferg {erg cm$^{-2}$ s$^{-1}$}

\def \gray {$\gamma$-ray }

\def \pks {\hbox{PKS~1830-211~}}
\def \agile {AGILE}

\def \igr {INTEGRAL}
\def \swi {{\it Swift}}

\usepackage{subfigure}
\begin{document}

   \title{The remarkable \gray activity in the gravitationally lensed
     blazar PKS 1830-211 }

   \author{I. Donnarumma\altaffilmark{1,*},
 A. De Rosa\altaffilmark{1}, V. Vittorini\altaffilmark{1},
 H. R. Miller\altaffilmark{2}, L. \v{C}. Popovi\'{c}\altaffilmark{3,4}, S. Simi\'c\altaffilmark{4,5} 
 M. Tavani\altaffilmark{1,6}, J. Eggen\altaffilmark{2}, J. Maune\altaffilmark{2}, E. Kuulkers\altaffilmark{7},
 E. Striani\altaffilmark{1,6}, S. Vercellone\altaffilmark{8},
 G. Pucella\altaffilmark{9}, F. Verrecchia\altaffilmark{10},
 C. Pittori\altaffilmark{10}, P. Giommi\altaffilmark{10}, L. Pacciani\altaffilmark{1},
 G. Barbiellini\altaffilmark{11}, A. Bulgarelli\altaffilmark{12},
 P. W. Cattaneo\altaffilmark{13}, A. W. Chen\altaffilmark{14}, E. Costa\altaffilmark{1}, E. Del
 Monte\altaffilmark{1}, Y. Evangelista\altaffilmark{1},
 M. Feroci\altaffilmark{1}, F. Fuschino\altaffilmark{12}, F. Gianotti\altaffilmark{12},
 A. Giuliani\altaffilmark{14}, M. Giusti\altaffilmark{1}, F. Lazzarotto\altaffilmark{1},
 F. Longo\altaffilmark{11}, F. Lucarelli\altaffilmark{10},
 A. Pellizzoni\altaffilmark{15}, G. Piano\altaffilmark{1}, P. Soffitta\altaffilmark{1},
 M. Trifoglio\altaffilmark{12}, A. Trois\altaffilmark{15}
}

\altaffiltext{1}{INAF/IASF--Roma, Via del Fosso del Cavaliere 100,
  I-00133 Roma, Italy}

\altaffiltext{2}{Dept. of Physics \& Astronomy
    Georgia State University, GA 30303-3083 U.S.A.}

\altaffiltext{3}{Astronomical Observatory, Volgina 7, 11160, Belgrade 74, Serbia}
\altaffiltext{4}{Isaac Newton Institute of Chile, Yugoslavia Branch}
\altaffiltext{5}{Faculty of Science, Department of Physics, University of Kragujevac, Radoja Domanovića 12, 34000 Kragujevac, Serbia}
\altaffiltext{6}{Dip. di Fisica, Univ. ``Tor Vergata'', Via della Ricerca Scientifica 1,I-00133 Roma, Italy}
\altaffiltext{7}{European Space Astronomy Centre, SRE-O, Villanueva de la Ca\~nada (Madrid), Spain}
\altaffiltext{8}{INAF/IASF Palermo Via Ugo La Malfa 153, 90146 Palermo, Italy}
\altaffiltext{9}{ENEA--Frascati, Via E. Fermi 45, I-00044 Frascati (Roma), Italy}
\altaffiltext{10}{ASI--ASDC, Via G. Galilei, I-00044 Frascati (Roma), Italy}
\altaffiltext{11}{Dip. di Fisica and INFN Trieste, Via Valerio 2, I-34127 Trieste, Italy}
\altaffiltext{12} {INAF/IASF--Bologna, Via Gobetti 101, I-40129 Bologna, Italy}
\altaffiltext{13} {INFN--Pavia, Via Bassi 6, I-27100 Pavia, Italy}
\altaffiltext{14}{INAF/IASF--Milano, Via E.~Bassini 15, I-20133 Milano, Italy}
\altaffiltext{15}{INAF-Osservatorio Astronomico di Cagliari, localit Poggio dei Pini, strada 54, I-09012 Capoterra, Italy}

\altaffiltext{*}{AGILE Team Corresponding Author: I. Donnarumma, immacolata.donnarumma@iasf-roma.inaf.it}

\slugcomment{accepted for publication in  ApJL}
\begin{abstract}

We report the extraordinary \gray activity ($E>100$ MeV) of the
gravitationally lensed blazar PKS 1830-211 ($z=2.507$) detected by
AGILE between October and November 2010. The source experienced on October 14 a flux
increase of a factor of $\sim$ 12 with respect to its average value and kept
brightest at this flux level ($\sim 500\times 10^{-8}$ \phcmsec) for about 4 days. The 1-month
\gray light curve across the flare showed a mean flux F($E>100$ MeV)= $200\times 10^{-8 }$ \phcmsec, which resulted in an enhancement
by a factor of 4 with respect to the average value. 
Following the $\gamma$-ray flare, the source was observed in NIR-Optical
energy bands at the Cerro Tololo Inter-American Observatory and in X-rays by \swi{}/XRT and \igr/IBIS. The main result of
these multifrequency observations is that the large variability observed in
$\gamma$-rays has not a significant counterpart at lower frequencies:
no variation greater than a factor of \textbf{$\sim 1.5$} resulted in NIR
and X-ray energy bands. \pks is then a good ``\gray only 
flaring'' blazar showing substantial variability only above 10-100 MeV. 
We discuss the theoretical implications of our findings.

\end{abstract}

\shorttitle{Observations of PKS 1830-211 in October 2010}
\shortauthors{I.Donnarumma}


%

\section{Introduction}

PKS 1830-211 is a high redshift blazar ($z=2.507$, Lidman et al. 1999)
gravitationally lensed by a spiral galaxy at $z=0.886$ (Wiklind \& Combes
1996), as showed by  the two radio lobes located 1'' apart from each other
(Lovell et al. 1998). The lensed counterparts were observed also in near
infrared (NIR) and
optical energy bands by Hubble Space Telescope and Gemini observatories (Courbin et al. 2002). In
X-rays the source was observed by both XMM-Newton and Chandra allowing to
study in details the complicated soft X-ray behaviour, probably due to absorption (Dai et al. 2008).
Moreover, the source presents a soft \gray energy spectrum ($\Gamma= 2.56$)
as detected by EGRET (Hartman et al. 1999) and recently confirmed by {\it
  Fermi} LAT (Abdo et al. 2010).

The modelling of the SED of this blazar (De Rosa et
al. 2005) led to the tentative classification of PKS 1830-211 as a MeV blazar
(the inverse Compton peak lies below 100 MeV, Sikora et al. 2002), with the
{\it   caveat } that the whole dataset was not simultaneous. In particular,
the SED was interpreted in the standard framework of a one-zone leptonic model
where the seed photons responsible for inverse Compton peak should be
originated within the torus surrounding the central engine.

The source has been extensively monitored in hard X-rays by \igr/IBIS on 7-year long period, during which no significant
variability has appeared on both short (daily) and longer (monthly)
timescales (Zhang et al. 2008).

Since its launch, \agile{} (Tavani et al. 2009) detected a \gray flare (F($E>100$
MeV)$=(160\pm 50) \times 10^{-8}$ ph cm$^{-2}$ s$^{-1}$) of PKS
1830-211 in October 2009 (Striani et al. 2009) which resulted in an
enhancement  by a factor of 3 with respect to the 1-week average flux detected before the flare (F($E>100$ MeV)$=60\times 10^{-8}$ photons cm$^{-2}$ s$^{-1}$).

One year later, {\it Fermi} LAT detected a significant enhancement of the \gray
emission F($E>100$ MeV)$=(520\pm 110)\times 10^{-8}$ photons cm$^{-2}$ s$^{-1}$  and
 F($E>100$ MeV)$= (1400\pm500)\times 10^{-8}$ photons cm$^{-2}$ s$^{-1}$ on 1-day and
 6-hr timescales, respectively (Ciprini et al. 2010), resulting in an
 increase of a factor of about  12 and 35 with respect to the average value
 reported by Abdo et al. 2010. Moreover, AGILE
detected a prolonged \gray activity two days after the flare reported by
{\it Fermi} (Donnarumma et al. 2010).

In this letter, we present the results obtained by means of a multifrequency
monitoring of PKS 1830-211 from NIR to \gray energy bands. In detail,
we describe the multi-$\lambda$ data analysis and a possible interpretation of
the SED of this object to reconcile the average \gray activity with the \gray
enhancement. We adopted a $\Lambda$-CDM cosmology with the following
parameters $h=0.71$, $\Omega_{m}=0.27$, $\Omega_{\Lambda}=0.73$.

\section{The multiwavelength campaign}

\subsection{\agile{} Observations}

AGILE/GRID data were analyzed using the Build-20 software. Well-reconstructed
\gray events were selected using the FM3.119 filter. All the events
collected during the passage in the South-Atlantic Anomaly were rejected. We
filtered out the Earth-albedo, rejecting photons coming from a circular region
of radius 80 degrees and centered on the Earth. A 1-month long \gray
light curve (2-day time-bin, see Fig. 1) across the \gray flare reported by
Fermi/LAT has been produced between 2010-10-08 00:00UT and 2010-11-08 00:00UT
(MJD 55477-55509) using the standard AGILE Maximum-Likelihood procedure. We
obtain a mean flux on the whole integration time
of F($E> 100$ MeV)=$(200\pm29)\times 10^{-8}$ photons cm$^{-2}$ s$^{-1}$ which is a
factor of $\sim 4$ higher than the flux reported in
the first {\it Fermi} catalog. We note that this $\gamma$-ray enhancement
represents a 5-$\sigma$ excess with respect to its value in steady state as
detected by AGILE (F($E>100$ MeV)$=(32\pm 8) \times 10^{-8}$ photons cm$^{-2}$
s$^{-1}$, Verrecchia et al. in preparation).

We extracted the \gray spectrum between $100 \rm MeV - 1
\rm GeV$, discarding the energy channels above 1 GeV due to the poor
statistics. The AGILE spectrum is well fitted by a power-law with a photon index of $2.05\pm0.17$  (all
the errors reported in this paper are at $1\sigma$, unless otherwise stated).

As shown in Fig.1, the source remained at a flux level $\sim 200 \times
10^{-8}$ photons cm$^{-2}$ s$^{-1}$ on a 1-month time scale. A clear enhancement
emerged on October 14 (MJD 55483) that lasted about 4 days.
The 2-day binned AGILE data showed that the source reached its maximum
  of  F($E> 100$ MeV)$\sim(500\pm130)\times 10^{-8}$ photons cm$^{-2}$ s$^{-1}$
  on October 15 (MJD $55484.75\pm 0.7$); the estimate of
  this epoch accounts for the bias introduced by the choice of the
  starting time (T$_{start}$) and binning adopted to obtain the light
  curve. This was done by operating 4 translations of T$_{start}$ (by
  $0$-, 0.5-, $1$-, $1.5$-day) in the light curve. 

We extracted the $\gamma$-ray spectrum also during the 4-day flare,
obtaining a power-law photon index $\Gamma=2.40\pm 0.28$.
Because of the poor statistics on this short duration, it was not possible to
infer any hardening/softening occurred during this flare, being the photon indices compatible within 1-$\sigma$.
In section 3, we will focus on the 1-month \gray enhancement and on
  the 4-day flare detected between October 14 and 18 (horizontal lines in
  Fig. 1).

\subsection{X-ray  observations}

\subsubsection{The {\it \igr} observations}

During the $\gamma$-ray flare of \pks in October 2010, \igr~  (Ubertini et
al. 2003) was monitoring the
Galactic bulge region (since \igr~ AO-3, Kuulkers et al. (2007) have been
observing this region regularly during all the visibility periods).
The monitoring across the $\gamma$-ray flare  (MJD $55479-55491$,
2010 October 10-22) did not allow a detection of
PKS 1830-211, then an upper limit of $\sim 3 \times 10^{-11}$ \ferg~ (it
accounts for the 15 degrees off-axis position of the source) was obtained on a net exposure of 200 ksec (October 14-18).
Since the hard X-ray average flux (20-40 keV) is 2.6 mCrab ($1.95 \times 10^{-11}$ \ferg, Bird et al. 2010),
we may exclude  variations greater than a factor of 1.5 in this energy band.
Therefore, in this paper we adopted the \igr/IBIS data
collected in the 4th IBIS survey (Bird et al. 2010) consisting of 7355
Science Windows (lasting about 2000 s each), performed from the
beginning of the Mission in November 2002 up to April 2008 and including
all available public and Core Programme data. The total on-source exposure is 4258 ks. IBIS/ISGRI images for each
available pointing were generated in various energy bands using the ISDC
off-line scientific analysis software OSA (Goldwurm et al. 2003) version
7.0. Then all images have been mosaicked to create significance map at
revolution level (each revolution lasting about 3 days) and for the all available
pointings. Count rates at the position of the source were extracted from
individual images in order to provide light curves in various energy bands;
from these light curves, we extracted and combined the average fluxes in order
to produce an average source spectrum (see Bird et al. 2010 for details).
The derived hard X-ray spectrum is well fitted by a power-law with
$\Gamma=1.56\pm0.14$, providing a flux of $1.4\times 10^{-11}$ erg cm$^{-2}$ s$^{-1}$ and $2.7\times
10^{-11}$ erg cm$^{-2}$ s$^{-1}$ in 20-40 keV and 40-100 keV, respectively.
It is worth noting that this spectrum represents the best coverage of PKS 1830-211 in the hard X-rays.

\subsubsection{The \swi{} observations}
Following the \gray flare, \swi{}/XRT (Gehrels et al. 2004) pointed at \pks
twelve times, between 2010 October 15--27 (MJD 55484-55496) with net exposures ranging between 1000 and 4000 s.
The XRT level 1 data were processed with the standard procedures ({\texttt{XRTPIPELINE
v.0.12.2}).
We selected photon counting (PC) mode data with the standard 0--12 grade selection, and
source events were extracted in a circular region of  aperture $\sim 40''$,
the background was estimated in different same sized circular regions far from the source.
We created response matrices through the \texttt{xrtmkarf} task.
As for spectra with very few counts, we applied the Cash statistics and we
rebinned the remaining spectra in order to have at least 30 counts per bin,
excluding from the fit the channels with energies below 0.3 keV and above 10 keV.
We fitted each spectrum, with a continuum  power--law absorbed both with a Galactic
column density ($N^{\rm Gal}_{\rm H}$=2.6$\times$10$^{21}$ cm$^{-2}$ Stark et
al. 1992) and an
additional absorber located at a redshift z=0.886. Due to the short exposure for each
observation, the fitting procedure resulted in a degeneracy between $N^{\rm z}_{\rm H}$
and $\Gamma$. The values of $N^{\rm z}_{\rm H}$ we found combining all the
spectra together are in full agreement
with the previous measurement obtained with Chandra observations, therefore we
decided to fix $N^{\rm z}_{\rm H}$ to this value, i.e. 1.94$\times$10$^{22}$ cm$^{-2}$.
All data sets are well reproduced by this model, with $\chi^2$/dof ranging between 0.9 and 1.1. The best fit values of the photon index  range between
$\Gamma$=1.0$\pm$0.3-1.4$\pm$0.2 with absorbed fluxes in 0.3--10 keV between
0.9$^{+0.1}_{-0.1}$ and 1.2$^{+0.1}_{-0.2}\times $10$^{-11}$ erg cm$^{-2}$
s$^{-1}$, all consistent within 1-$\sigma$ given the large uncertainties. On
2010 October 15, the day overlapping the gamma-ray flare, the absorbed flux in 0.3--10 keV is  1.1$^{+0.2}_{-0.2}\times $10$^{-11}$ erg cm$^{-2}$ s$^{-1}$ at 90\% confidence level.
We note that these fluxes are well consistent with the ones reported by De
Rosa et al. (2005), ruling out any significant variation occurred during the
high \gray activity (Fig. 1). On the time interval of the gamma-ray enhancement (15--18
  oct. 2010), the full band detected count rates are
  between 0.095$^{+0.006}_{-0.006}$ and 0.148$^{+0.009}_{-0.009}$ counts/s, showing a variation amplitude that is less than a factor of 1.6.

\subsection{Optical monitoring:  SMARTs Consortium}
Optical and infrared data on PKS 1830-211 were obtained using the 1.3m
telescope at the Cerro Tololo Inter-American Observatory under the
Small and Moderate Aperture Research Telescope System (SMARTS) program. We
obtained simultaneous data in the optical $R$ and infrared $J$ bands using the
ANDICAM instrument, allowing the acquisition of data from 0.4 to 2.2
microns. Our dataset consists of 8 $R$-band and 17 $J$-band images, taken
between 2010 October 16 and 27 (MJD 55485-55496). 
Images were taken approximately every three nights, with exposure times of 400
to 600 seconds in $R$, and 90 to 600 seconds in $J$. Both the optical and IR
data were flat-fielded, bias-subtracted, and over-scan corrected using
standard IRAF packages. Single exposures were not sufficient to
detect the object in either waveband, nor were summations of the images using
standard IRAF packages. We estimate a limiting magnitude of 21.5 in
  $R$ and 17.6 in $J$, based on the combined images and the magnitudes of
  field stars acquired from the 2MASS. The observations performed during the 4-day gamma-ray flare resulted in a
  limiting magnitude 17.6 and 19 in $J$ and $R$, respectively. The $J$ magnitude
  limit during the flare period remained unchanged from the longer summation
  due the sky limit already reached by the shorter exposure time. 
  We note that our limit magnitude could be contaminated by the star S1 at
  $1"$ from PKS 1830-211 (Meylan et al. 2005) unresolved by
  SMARTs. Neverthless, no enhancement at or above two standard deviations of the background (sky) variance was
  detected within $2"$ of the position of PKS 1830-211. Therefore, by
  comparing our limiting magnitude in J with the photometric data of PKS
  1830-211 ($J=18.04$), we exclude variation $\geq 1.5$ in $J$-band.

\section{Discussion}

Here we discuss the SED modelling of the blazar PKS 1830-211 during the high \gray state observed between
October-November 2010. Multifrequency observations were activated to follow
the \gray flare covering NIR-Optical (SMARTs), soft and hard
X-ray energy bands (\swi{}/XRT and \igr/ISGRI, respectively).
The main results obtained during this multifrequency campaign are that
the high and unusual activity recorded in $\gamma$-rays seems to have no
significant counterpart at lower frequencies. In detail, the simultaneous
NIR-optical, soft and hard X-ray emissions of this source did not follow the significant changes
observed in $\gamma$-rays.
As for the NIR and optical emissions, we derived only upper limits
for this source that is indeed vary faint especially in optical energy bands
(Courbin et al. 2002). Neverthless, by comparing the derived upper limits
in $J$-band (red downward arrow in Fig. 2) with
the NIR data reported by Courbin et al. 2002 and Meylan et al. 2005, we can exclude variation in
the thermal + non-thermal components greater than a factor of $\sim 1.5$. The
soft and hard X-ray emissions also exclude variation greater than a factor of
1.6 and 1.5, respectively. It is worth noting
that the observed variations of the SED rule out the hypothesis that the
\gray emission was connected to macrolensing, since its effects
would be energy-independent. On the other hand, the chromaticism of
  the SED variability may suggest that microlensing from stars in the
  lensing galaxy may cause the observed gamma-ray variability (see e.g. Torres
  et al. 2003). In order to investigate this possibility, we simulated the transition of a source with different
  dimensions (Jovanovi\'c et al. 2008) across the microlensing
  magnification pattern for the lens system (taking convergence
  $k=0.158$ and shear $\gamma=0.096$ as given in Winn et al. 2002). By
  assuming a gamma-ray emitting region of $10^{15}$ cm (observer frame), we found an amplification
  by a factor of tens on time scale of  years. Even worst
  is the case of the gamma-ray flare observed by Fermi on 6-hr time scale, 
  which is a factor of 14 higher than the steady state, requiring high
  amplification in very short time. Shorter time-scales
  of microlensing can be obtained with very compact source (order of $10^{13}$ cm),
  but we cannot further reduce the gamma-ray emitting region otherwise it
  would result in an increase in gamma-ray opacity due to pair
  production. Therefore, we found that the observed variability 
is more likely intrinsic to this blazar.

We complemented the low frequency data with the non-simultaneous {\it
  Planck} observations (30, 100, 217 GHz) taken from the Early Release Compact
Source Catalogue (ERCSC) extracted from the first all-sky survey of Planck\footnote{\url{http://www.sciops.esa.int/index.php?project=planck\&page=Planck_Legacy_Archive}}.
In Fig. 2 we report all the observed multifrequency data and the SED modelling taking into account the acromatic magnification due to the gravitational lensing, assumed to be one order of
magnitude (De Rosa et al. 2005).

We model three \gray states of the source. The first one is the
average state reported by De Rosa et
al. 2005 and consistent with the average $\gamma$-ray state reported in Abdo et
al. 2010 ($\sim40\times 10^{-8}$ ph cm$^{-2}$ s$^{-1}$) (see black points in
Fig. 2);
the second is the 1-month enhancement showing a mean flux of $\sim$ 200
$\times$ 10$^{-8}$ ph cm$^{-2}$ s$^{-1}$ (blue points in Fig. 2); the third is the peak of the \gray emission
(October 14-18; hereafter, ``flare'') which
is a factor of $\sim 2-3$ higher than the 1-month average flux (red points in
Fig. 2).
We note that the multifrequency behavior of the source, i.e. the
lack of correlated variability between the low (NIR-optical bands, X-rays) and
high energy ($\gamma$-rays) portions of the
SEDs, disfavors the one-zone leptonic model for this event.
We account for this evidence, assuming two relativistic electron
populations with the energy density described by a  broken power law in the Lorentz factor domain ($\gamma$) with break equal to $\gamma_{b}$:
\begin{equation}
n_{e}(\gamma)=\frac{K\gamma_{b}^{-1}}{(\gamma
/\gamma_{b})^{a_{l}}+(\gamma /\gamma_{b})^{a_{h}}}\,
\end{equation}
where $a_{l}$ and $a_{h}$ are the pre-- and post--break electron distribution spectral indices,
respectively. We assumed that the two electron populations contained a random magnetic
field $B$. In Table 1 we summarize the values of the model parameters for the
two components.

We estimated the contributions of both electron populations to sychrotron
emission and  to synchrotron self
Compton (SSC) plus inverse Compton from external seed photons (EC) originating
in a dusty torus, in the accretion
disk and in the Broad Line Region (BLR) (see Sikora et al. 2002, Vittorini et al. 2009).
In detail, we assume: 1) the emission from a torus
with temperature $T= 10^3$ K distant $1.26\times 10^{19}$ cm from the central engine 2); the direct emission from the
disk with $T = 1.5 \times 10^{4}$ K and luminosity $L_{disk}=6\times 10^{45}$
erg s$^{-1}$  distant $5\times 10^{16}$ cm from the 2nd population; 3) the emission from the BLR distant R$_{BLR}=5\times 10^{17}$
cm from the central engine and reprocessing 10\% of the disk luminosity.

In this scenario, the average \gray state is produced through EC
on dusty torus, disk and BLR seed photons mainly by
the first population, resulting in a jet
power $P^{1}_{jet}=10^{46}$ \ergsec (black line in Fig. 2).
The 1-month \gray enhancement, a factor of 4 higher than the
average flux, has required the additional contribution of the second electron population, which
is characterized by a smaller size, higher electrons density and higher
$\gamma_{b}$ with respect to the 1st one (Table 1). On the basis of these assumptions, the \gray
enhancement (blue points in Fig.2) is then produced by EC of the 2nd
population on the photon density field while moving inside the BLR  with
 $\Gamma =15$, transporting a jet power $P^{2}_{jet}=3\times10^{46}$ \ergsec. This modelling is in agreement with the source fading at its
 average flux level when the blob is moving outside the BLR ($\Delta t\sim R_{BLR}(1+z)/c\delta \gtrsim 30$ days).
 Following this description, the synchrotron and IC emissions are dominated by the 1st population, thus accounting for the moderate variation inferred by the
NIR-optical and X-ray observations.
On the other hand the ``flare'', which is a factor of
12 greater than the average flux and a factor of $\sim 2-3$ greater than the 1-month activity,
requires a ``local'' enhancement of the photon
density field by a factor of 3 (red line in Fig. 2), likely due to a blob-cloud
interaction (Araudo et al. 2010). In this case we derived a jet power of $P^{flare}_{jet}=6\times10^{46}$ \ergsec.
The lack of correlated variabilities in optical, X- and $\gamma$-rays during 
the flare prevented us to explain this event as associated with
changes of the electron population in terms of injection/acceleration
mechanisms.

We underline that the \gray behavior of PKS 1830-211 recorded by AGILE and
{\it Fermi}/LAT between October-November 2010 is rare as probed by the
lack of similar variability since 2007.
Similarly, no hard X-ray variability has been observed during the 7-year
monitoring by \igr/IBIS. In conclusion, we can attribute the properties of
\pks to the ones of ``\gray only flaring'' blazars showing substantial
variations especially in the \gray energy range above 10-100 MeV.
These evidences strongly support our
interpretation of a ``steady'' electron population, filling the jet below the
BLR, that is responsible for the average $\gamma$-ray emission recorded by
both EGRET and {\it Fermi}/LAT, more than ten years apart.

It is worth noting that the inspection of the \gray light curve on a longer
period across the October-November enhancement, did not show any evidence of
the time delay between the emissions of the two lensed images A and B as measured in
the radio maps ($26^{+5}_{-4}$ days, Lovell et al. 1998). We can assume
the lack of the delay between A and B only if the flux ratio of the 2
components is $\sim 1$. If it was lower than 1, we are prevented to draw any
conclusion because the emission would be occurred below the AGILE
sensitivity level. We also note that microlensing could play a role in
explaining the possible lack of echo in the gamma-ray light curve. 
 In fact, in some lensed quasars (Blackburne et al. 2006), different flux ratios in the lensed images were detected between the optical and X-rays. 
This dependency of the flux ratio on the energy has been interpreted as due to
microlensing, thus justifying different amplifications as a function of the
emitting region size (see also Jovanovi\'c et al. 2008).  On the basis of these evidences, we can consider that the lack of the gamma-ray echo  may be due to  different flux ratios between the 2 components in radio and gamma-ray energy bands, i.e. it might be that  one component (due to the flux anomaly) has negligible contribution in gamma-rays. Consequently intrinsic variation in combination with microlensing may be also considered.

\section{Acknowledgements}
We thank the anonymous referee for helpful comments and suggestions. We warmly thank the INTEGRAL/IBIS survey Team for providing us the 7-year
hard X-ray spectrum of PKS 1830-211.
Research partially supported by ASI grant no. I/040/10/0. L.C.P and S.S. are
supported by Serbian Ministry of Science (project 176001).




\newpage

\begin{figure}[t]
\centering
\includegraphics[angle=0,scale=0.7]{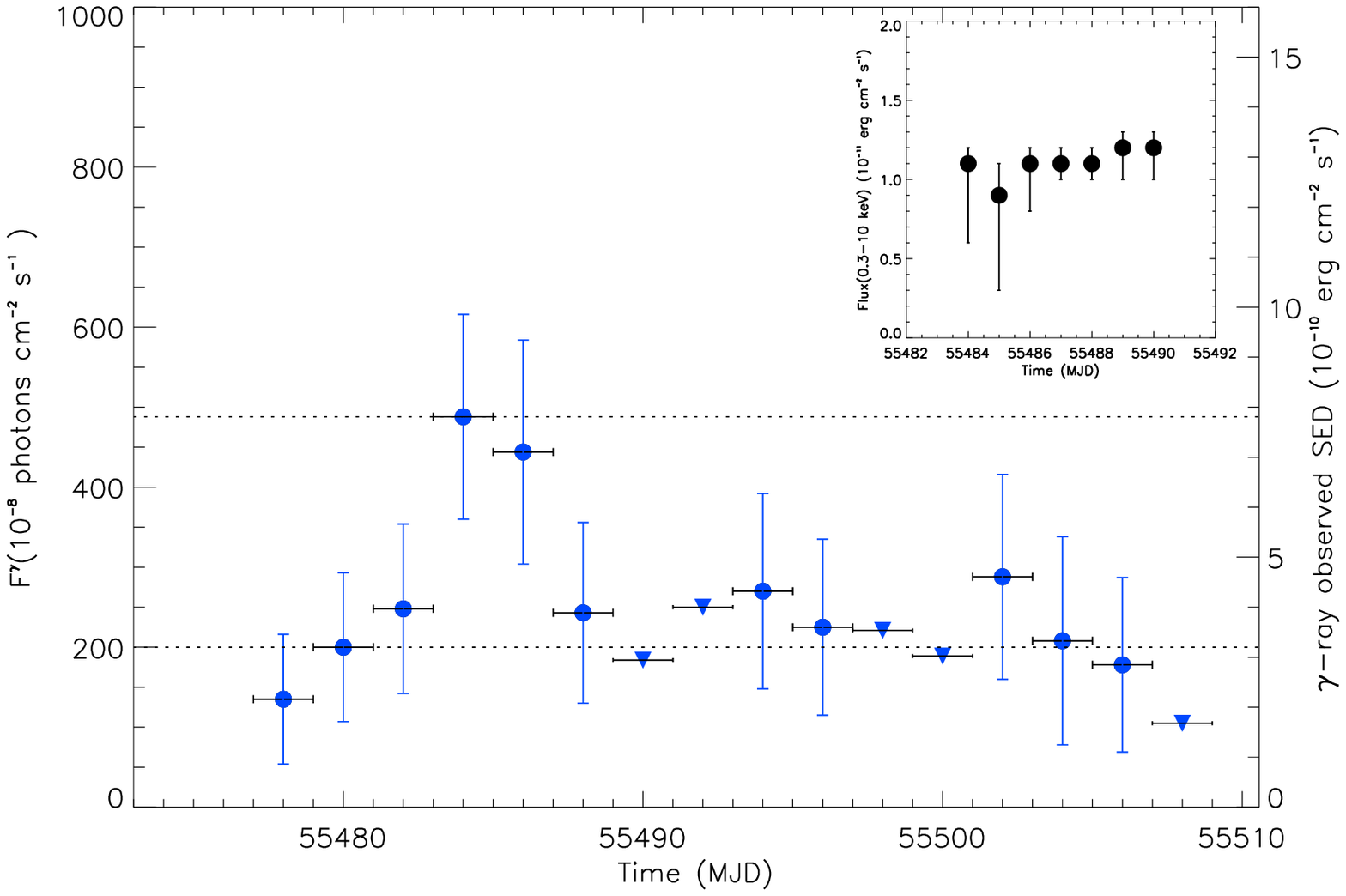}
\caption{The 1-month E$> 100$ MeV light curve (2-day time-bin) of PKS 1830-211
  (2010 October 8 - 2010 November 8). The two horizontal lines represent the
  1-month and the 4-day \gray enhancements. On the right $y$-axis we report the
  observed \gray SED. The downward triangles indicate the
  2-$\sigma$ \gray upper limits. The plot insert shows the soft X-ray
    light curve.}
\end{figure}

\begin{figure}[ht]
\centering
\includegraphics[angle=0,scale=0.7]{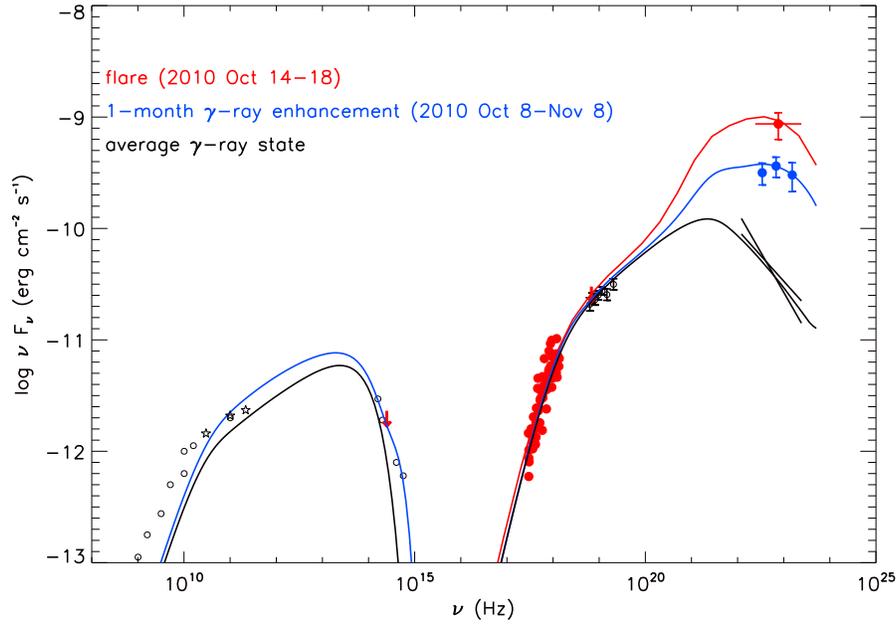}
\caption{The observed SED of PKS 1830-211 for the \gray flare (red filled
  circles), for the 1-month enhancement (blue filled
  circles) and for the average state (black points). In detail, the black open circles are the non-simultaneous data from radio to hard
  X-rays (Pramesh Rao \& Subrahmanyan 1998, Courbin et al. 2002, 7-year
  INTEGRAL survey, respectively). We report as open stars Planck data taken
  from the ERCSC of the first all-sky survey of Planck. The red downward
  arrows represent the near-IR {\it SMARTS} and
  the INTEGRAL/IBIS (20-40 keV) upper limits. Solid lines represent
  the SED models for the three \gray states (black for the average (EGRET),
  blue for the 1-month, red for the flare). The models were magnified by a factor of ten for the lensing.}
\end{figure}
\nocite{*}


\begin{table}
\begin{center}
\caption{Model Parameters for 2010 \gray activity
of 1830.}

 \small \noindent
\bigskip
\begin{tabular}[t]{|c|c|c|c|c|c|c|c|c|c|}
  \hline
  \bf{population} & $\bf{\Gamma}$ & \bf{B(Gauss)} & \bf{R(cm)} & $\bf{K(cm^{-3})}$ &
  $\bf{\gamma_b}$ & $\bf{\gamma_{min}}$ & $\bf{a_l}$ & $\bf{a_h}$ &
  $\bf{\delta}$ \\
  \hline
   1st  & 10 & 0.7 & $8\times10^{16}$ & 100 & 100 & 35 & 2.0 &2.6 & 16 \\

  \hline
   2nd  & 15 & 0.4 & $3\times10^{16}$ & 150 & 500 & 60 & 2.0 & 3.4 & 20 \\
\hline
\end{tabular}
\end{center}
\end{table}

\end{document}